\documentstyle[epsfig,12pt,dina4]{article}
\input{epsf}                    % zum Einbinden von eps-files
\def\VYP#1#2#3{ #1: #3 (#2)}  % Volume Year Page, Plenum format
\def\NP#1#2#3{{\it Nucl.\ Phys.\ }\VYP{#1}{#2}{#3} }
\def\PL#1#2#3{{\it Phys.\ Lett.\ }\VYP{#1}{#2}{#3} }
\def\PR#1#2#3{{\it Phys.\ Rev.\ }\VYP{#1}{#2}{#3} }
\def\PRL#1#2#3{{\it Phys.\ Rev.\ Lett.\ }\VYP{#1}{#2}{#3} }

\def\ZP#1#2#3{{\it Z.\ Phys.\ }\VYP{#1}{#2}{#3} }

\begin{document}
\renewcommand{\thefootnote}{\fnsymbol{footnote}}

\baselineskip=16pt
%=======================================================================

\begin{center}

{\Large\bf
Multifragmentation 
\vspace*{0.3cm}

in Relativistic Heavy Ion Reactions\footnote{
Lecture Notes of NATO Advanced Study Institute, Dronten, The Netherlands, 
1996,
to appear in {\it Correlations and Clustering Phenomena in Subatomic 
Physics}, edited by M.N. Harakeh, O. Scholten, and J.K. Koch,
Plenum Publishing Corporation.}
}
\vspace*{0.8cm}

W. Trautmann

Gesellschaft f\"ur Schwerionenforschung mbH

D-64291 Darmstadt, Germany

\vspace{1.0cm}

{\bf Abstract}
\vspace{0.3cm}

\end{center}

Multifragmentation is the dominant decay mode of heavy nuclear systems 
with excitation energies in the vicinity of their binding energies.
It explores the partition space associated with the number 
of nucleonic constituents and it is characterized by a multiple production
of nuclear fragments with intermediate mass.

Reactions at relativistic bombarding energies, exceeding
several hundreds of MeV per nucleon, have been found 
very efficient in creating such highly excited systems. 
Peripheral collisions of heavy symmetric systems or more central
collisions of mass asymmetric systems produce spectator
nuclei with properties indicating a high degree of equilibration.
The observed decay patterns are well described by 
statistical multifragmentation models.

The present experimental and theoretical studies are particularly
motivated by the fact that
multifragmentation is being considered a possible manifestation of the
liquid-gas phase transition in finite nuclear systems.
From the simultaneous measurement of the temperature and of the energy content
of excited spectator systems a caloric curve of nuclei has been obtained.
The characteristic S-shaped behavior resembles that of ordinary liquids. 

Signatures of critical phenomena
in finite nuclear systems are searched for in multifragmentation
data. These studies, supported by the success of percolation in
reproducing the experimental mass or charge correlations, concentrate on
the fluctuations observed in these observables. Attempts have been made
to deduce critical-point exponents associated with multifragmentation.

\newpage

\section{Introduction}
\label{Sec_1}

The hope to establish a link to the liquid-gas phase transition in
nuclear matter has been a major motivation for the search for and the
study of multi-fragment decays of heavy nuclei in recent years
\cite{siemens,huefner}.
Multifragmentation was predicted to be the dominant decay mode at excitation
energies near the binding energy of nuclei of about 8 MeV per nucleon
and at densities below the saturation density of nuclear matter
\cite{gross1,bondorf2}.
These conditions of high excitation and low density 
coincide with the liquid-gas coexistence region
as predicted for nuclear matter from the Van-der-Waals type range 
dependence of the nuclear forces \cite{jaqaman,brack}. It was also
suggested early on that this region may be explored during the
later stages of energetic nuclear reactions \cite{bertsch}. 

The experimental study of the nuclear liquid-gas phase transition
in finite nuclei faces several serious difficulties related to the fact
that excited nuclei are composed of a small number of constituents,
that they are charged, and that there is no external pressure to
counteract the internal pressure of the system at a given equilibrium 
condition \cite{stocker}.
Finite pressures may be maintained only dynamically and for very short 
periods of time during the disintegration process.
There is also no heat bath available which 
would allow to predetermine the temperature of the system in order 
to measure its response to it.

In spite of these difficulties, stimulating new results have been 
presented very recently. They
suggest that signals of the nuclear liquid-gas phase transition
may be revealed by studying reactions of finite nuclei.
From the simultaneous measurement of the temperature and the excitation 
energy for excited projectile spectators 
in $^{197}$Au + $^{197}$Au collisions 
at 600 MeV per nucleon a caloric curve of nuclei has 
been obtained \cite{pocho1}. It exhibits a typical 
S-shaped behavior, reminiscent of 
first-order phase transitions in macroscopic systems.
For the $^{197}$Au on C reaction at 1.0 GeV per nucleon,
the EOS collaboration has reported values of critical-point exponents
which were derived from the correlations and fluctuations of the
fragment sizes \cite{gilkes}. In both cases, the data providing the
basis for the analysis were obtained from the fragmentation of heavy
projectiles at relativistic energies in the range of up to 
about 1 GeV per nucleon. The decay properties of
spectator nuclei produced in these reactions indicate
that a high degree of equilibrium has been reached.
This is a prerequisite for the 
study of the thermodynamic behavior of highly excited nuclear matter
and makes these reactions rather attractive for 
this purpose.

\begin{figure}[ttb]
   \centerline{\epsfig{file=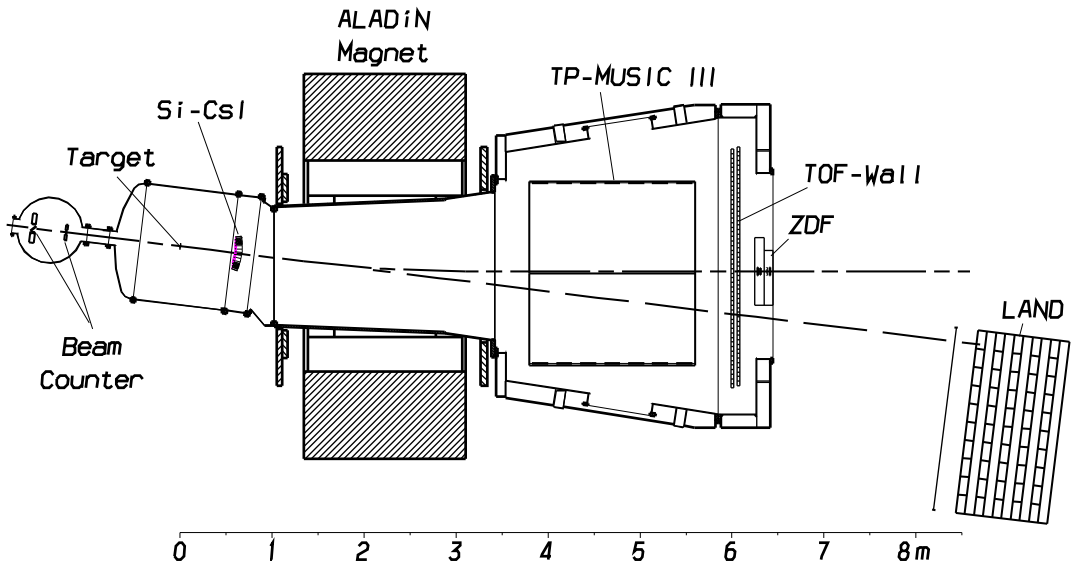,width=15cm,
   bbllx=80,bblly=300,bburx=480,bbury=520}}
        \caption[]{\it\small
Cross sectional view of the ALADIN facility in the configuration of the
1993 experiment. The beam enters from the
left and is monitored by two beam detectors before reaching the target.
Projectile fragments entering into the acceptance of the magnet are
tracked and identified in the TP-MUSIC III detector and in the
time-of-flight (TOF) wall. The Central Plastic detector (ZDF) covers the hole
in the TOF wall at the exit for the beam.
Fragments and particles emitted in forward
directions outside the magnet acceptance and up to
$\theta_{lab} = 16^{\circ}$ are detected in the Si-CsI array.
Neutrons emitted in directions close to $\theta_{lab} = 0^{\circ}$ 
are detected with the large-area neutron detector (LAND).
The dashed line indicates the direction
of the incident beam. The dash-dotted line represents the trajectory of
beam particles after they were deflected by an angle of 7.3$^{\circ}$
(from Ref. \cite{schuetti}).
        }
        \label{fig1}
\end{figure}

In the following notes, some main features of
multifragment-decays following heavy-ion reactions in the relativistic
regime of bombarding energies will be summarized. 
The experimental material will be mostly taken from the
work of the ALADIN collaboration, performed at the heavy-ion
synchrotron SIS of the GSI in Darmstadt \cite{schuetti,recent}. 
The techniques used to determine
the observables related to the liquid-gas phase transition will be briefly
described. It is not intended, however, to give a complete account of 
the present discussion initiated by these results. Besides the references
cited, the reader is referred to the proceedings of recent conferences
or workshops [13-16]
%\cite{hirschegg,bormio95,snowbird,acicast} 
during which the topic has been discussed within a wider
perspective. Further references on the subject of multifragmentation
in general may be found in \cite{moretto}.

\section{Experimental study of projectile decay}
\label{Sec_2}

The first observations of multi-fragment decays of heavy projectiles 
have been made by exposing nuclear emulsions
to the heavy ion beams \cite{jakob,fried}. 
This technique is still being used and, just recently,
has produced first results on the fragmentation of gold nuclei at the energy
of 10.6 GeV per nucleon, available from the AGS in Brookhaven
\cite{jain2,klmm2}. Another technique, employed in several studies,
is based on plastic nuclear track detectors which have high charge 
resolution for atomic numbers $Z \ge$ 6 \cite{rusch}.

\begin{figure}[ttb]
	\centerline{\epsfig{file=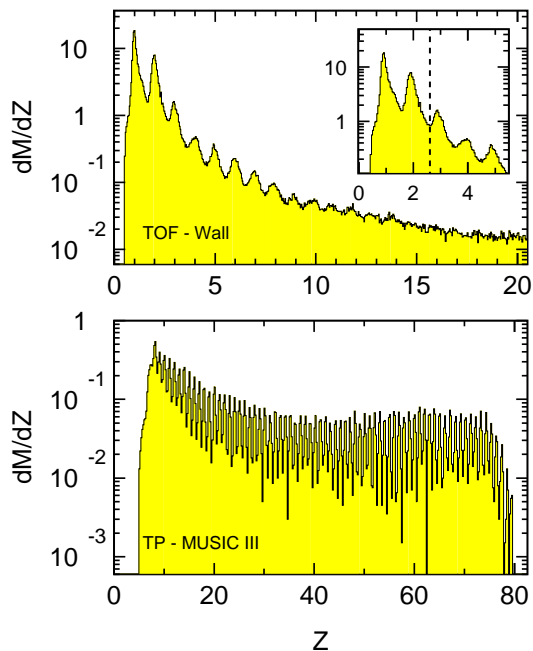,height=9cm}}
        \caption[]{\it\small
$Z$-identification spectra measured with the TOF wall
(top) and the TP-MUSIC (bottom) for the reaction
$^{197}$Au on $^{197}$Au at $E/A$ = 600 MeV.
The $Z$ information from the TP-MUSIC
was used to calibrate the response of the TOF wall in the
region $Z >$ 15.
Note that
the element yield at $Z >$ 65 is affected by the experimental trigger.\\
Insert: Low-$Z$ part of the TOF-wall spectrum.
The dashed line indicates the equivalent sharp cut which
was used for selecting fragments with $Z \ge$ 3 
(from Ref. \cite{schuetti}).
        }
        \label{fig2}
\end{figure}

More detailed investigations are possible with electronic detection devices
such as the ALADIN spectrometer at SIS \cite{schuetti,hubele1}. 
This detector system 
is built around A Large Acceptance DIpole magNet (ALADIN) with
the target in front of the magnetic-field gap
and the main detector systems behind it.
The configuration 
used in the 1993 experiment, a systematic study covering a wide
range of beams and targets, is shown in Fig.~1.
In this experiment, complete acceptance within the kinematic region
of projectile decay with good resolution was achieved.
The solid angle adjacent to the acceptance of the ALADIN magnet was covered
with 84 Si-CsI(Tl) telescopes in closely packed geometry.
Behind the magnet, the MUltiple Sampling Ionization Chamber (MUSIC)
served as the tracking
detector. The high charge resolution of this detector
permitted the identification of individual elements above
a threshold atomic number $Z \ge$ 8 (Fig.~2, bottom panel). 
Lighter fragments were tracked and identified by collecting
and amplifying
their ionization charges with proportional counters mounted in three
positions at the anode plane of the drift volume. 

\begin{figure}[ttb]
	\centerline{\epsfig{file=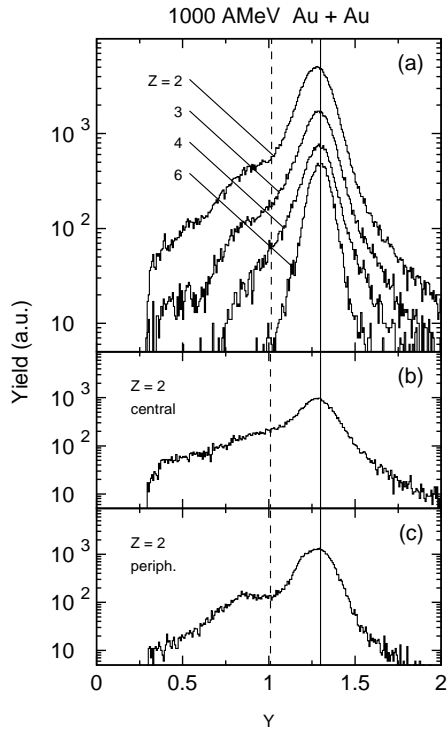,height=10cm}}
        \caption[]{\it\small
(a): Rapidity spectra measured in the reaction $^{197}$Au on $^{197}$Au at
$E/A$ = 1000 MeV for fragments with $Z$ = 2, 3, 4, and 6.
The solid and dashed lines indicate the measured most probable
rapidity $y =$ 1.32
of the light fragments and the condition $y \geq 0.75\cdot y_P$
adopted for fragments from the projectile spectator, respectively.\\
(b): Rapidity spectra of helium fragments, measured in central
collisions ($Z_{bound} \leq$ 30)
for the same reaction.\\
(c): Same as (b) for peripheral collisions ($Z_{bound} \geq$ 50)
(from Ref. \cite{schuetti}).
        }
        \label{fig3}
\end{figure}

The two-layered
time-of-flight (TOF) wall extended over 2.4 m in the horizontal and 
1.0 m in the vertical directions. 
Fragments with $Z \leq$ 15 were elementally resolved 
with the TOF wall detectors (Fig.~2, top panel).
The resolution of the time-of-flight measurement with respect to the
beam detectors positioned upstream was between 200 ps and 400 ps (FWHM),
depending on the fragment $Z$. It permitted the determination 
of the individual masses of the lighter products with $Z$ up to
about 12 from the momenta given by the tracking analysis.
In addition to the charged projectile fragments, neutrons emitted 
in directions close to 
$\Theta_{lab} = 0^{\circ}$ were measured with the Large-Area 
Neutron Detector (LAND) which was operated in a calorimetric mode.

The fragments from the decay of excited projectile spectators are
well localized in rapidity. This is illustrated in Fig.~3 where
rapidity spectra of light fragments from
the reaction $^{197}$Au on $^{197}$Au at
$E/A$ = 1000 MeV are shown. The distributions are
concentrated around a rapidity value very close to that of the 
projectile, $y_{P}$,
and become increasingly narrower with increasing
mass of the fragment. 
For the lighter fragments, the distributions 
extend into the mid-rapidity region. The widths and
shapes of the distributions also depend on the impact parameter, as
demonstrated for helium fragments in the two lower panels of Fig. 3.
The bump in the peripheral He spectrum, located at a rapidity $y$ between
0.8 and 0.9, and similar bumps observed for light fragments
up to $Z \approx$ 4 (Fig. 3, top) originate from mid-rapidity emission. 

\begin{figure}[tbh]
	\centerline{\epsfig{file=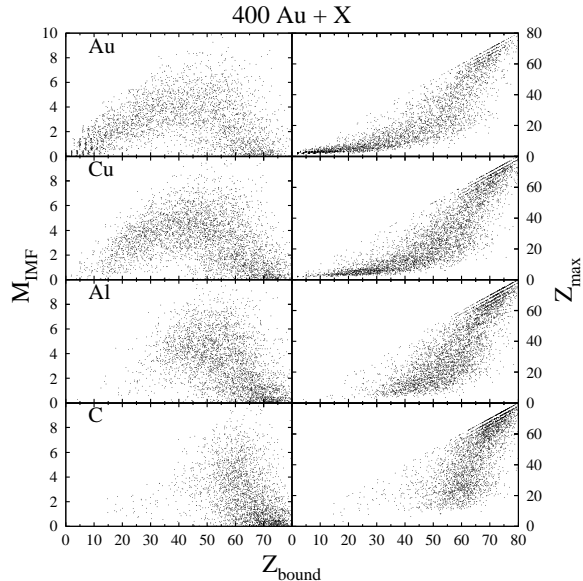,height=8cm}}
        \caption[]{\it\small
Multiplicity of intermediate-mass fragments (left-hand side) and atomic
number of the largest fragment (right-hand side) as a function of 
$Z_{bound}$ for the reaction $^{197}$Au on $^{197}$Au at $E/A$ = 400 MeV.
Random numbers, taken
from the interval [-0.5,0.5], were added to the integer values $M_{IMF}$
in order to preserve the intensity information in the scatter plot
(from Ref. \cite{kunze}).
        }
        \label{fig4}
\end{figure}

Based on these observations and on model studies, 
limits for the kinematic region of the projectile source
were chosen for the off-line analysis.
The condition $y \geq 0.75\cdot y_{P}$ 
was adopted for the fragment rapidities (cf. Fig. 3), and
upper limits in the laboratory angle were set which
took the observed invariance of the transverse fragment momenta
with bombarding energy into account.
These definitions permitted a comparison of data
measured at different bombarding energies on a quantitative level
\cite{schuetti}.

\section{Universality of spectator decay}
\label{Sec_3}

The decay of excited spectators exhibits universal features 
which become apparent in the
observed $Z_{bound}$ scaling of the measured charge correlations.
The quantity $Z_{bound}$ is defined
as the sum of the atomic numbers $Z_i$ of all projectile fragments
with $Z_i \geq$ 2. It represents the charge of the 
original spectator system
reduced by the number of hydrogen isotopes emitted during its decay.

\begin{figure}[tbh]
	\centerline{\epsfig{file=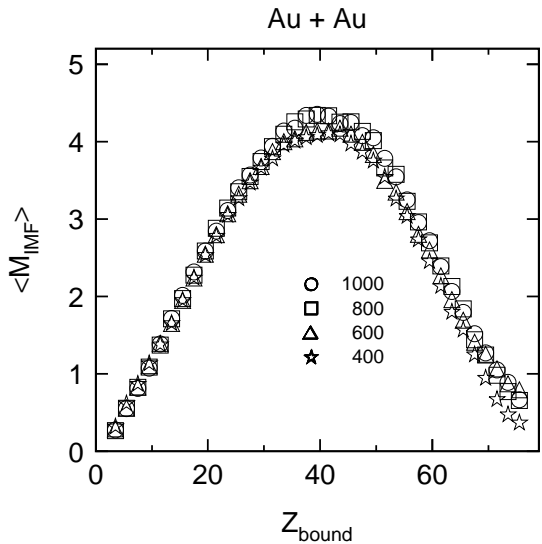,height=7.5cm}}
        \caption[]{\it\small
Mean multiplicity of intermediate-mass
fragments $\langle M_{IMF} \rangle$
as a function of $Z_{bound}$ for the reaction
$^{197}$Au on $^{197}$Au at $E/A$ = 400, 600, 800, and 1000 MeV
(from Ref. \cite{schuetti}).
        }
        \label{fig5}
\end{figure}

Scatter plots of two charge observables, the multiplicity $M_{IMF}$
of intermediate-mass fragments (IMFs, 3 $\le Z \le$ 30)
and the maximum fragment charge 
$Z_{max}$ within the event, are shown in Fig. 4 as a function of 
$Z_{bound}$ for $^{197}$Au projectiles at 400 MeV per nucleon
incident energy. The four rows of panels correspond to the results
obtained with the four targets C, Al, Cu, and Au. 
It follows from the geometric properties of heavy-ion reactions at these
energies that the mass (or charge) of the spectator, and therefore also
$Z_{bound}$, is closely related to the impact parameter. At large
$Z_{bound}$, the number of fragments is small and mainly one heavy
residue nucleus with $Z_{max} \approx Z_{bound}$ is produced. With
decreasing $Z_{bound}$ the number of fragments increases and,
correspondingly, $Z_{max}$ is considerably smaller than $Z_{bound}$.
Multi-fragment production dominates for impact parameters
corresponding to $Z_{bound} \approx$ 40. In
central collisions with the heavier targets, the region of small 
$Z_{bound}$ is strongly populated. 
Here both $M_{IMF}$ and $Z_{max}$ decrease, 
reflecting the smaller size of the spectators
produced in these collisions. This behavior was termed the rise and
fall of multi-fragment emission \cite{ogilvie}. 
On the side of the rise, at large $Z_{bound}$, the fragment production
is governed by the amount of deposited energy, whereas in the fall
region the limit of unconditional partitioning is approached 
\cite{hubele2}.

\begin{figure}[ttb]
	\centerline{\epsfig{file=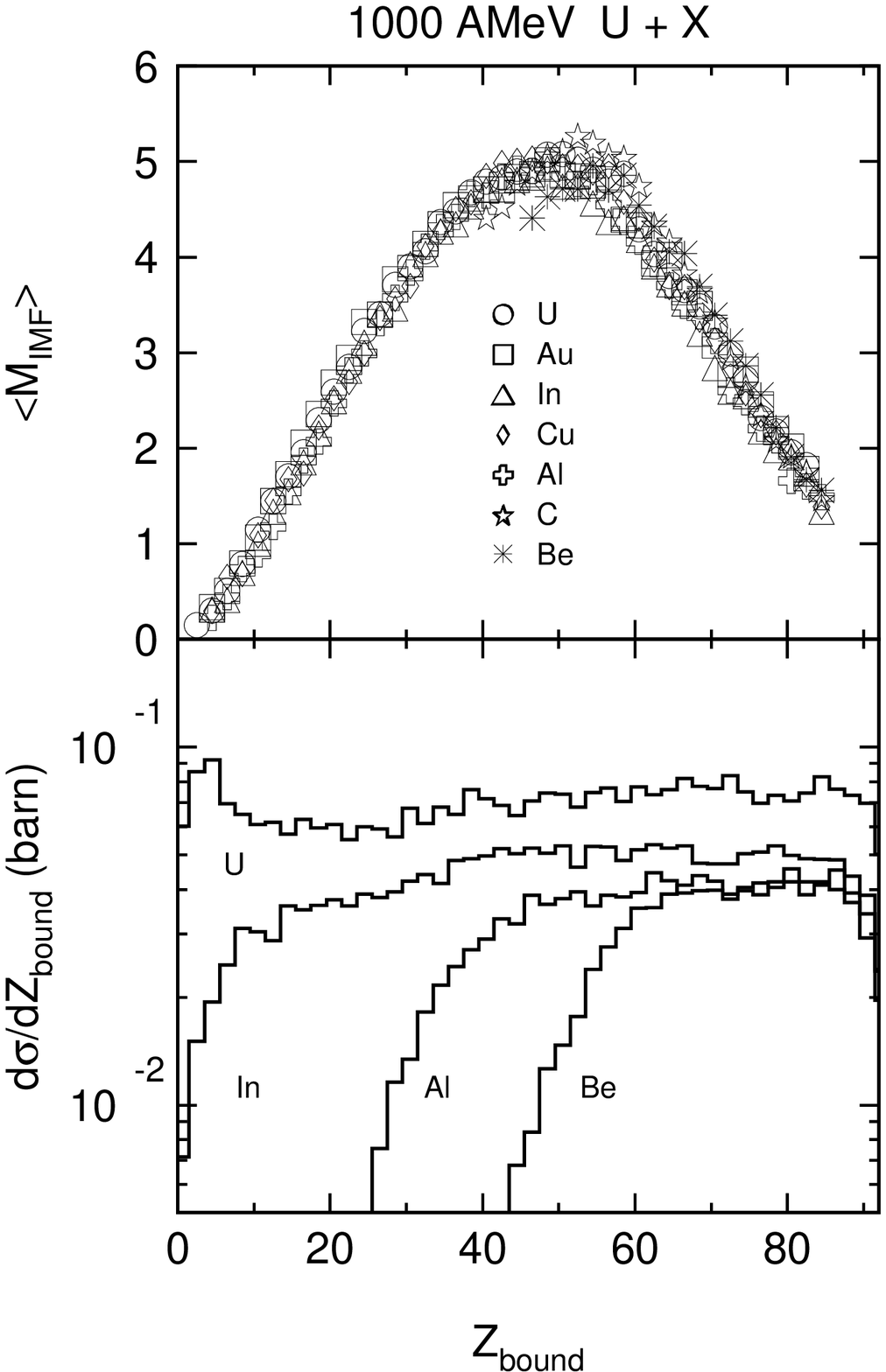,height=9cm}}
        \caption[]{\it\small
Top: Mean multiplicity of intermediate-mass
fragments $\langle M_{IMF} \rangle$ as a function of
$Z_{bound}$ for the reactions of $^{238}$U projectiles
at $E/A$ = 1000 MeV with the seven
targets Be, C, Al, Cu, In, Au, and U.

\noindent
Bottom:
Measured cross sections $d\sigma /dZ_{bound}$
for the reactions of $^{238}$U projectiles
at $E/A$ = 1000 MeV with the four targets
Be, Al, In, and U.
Note that the experimental trigger, for the case of uranium beams,
affected the cross sections
for $Z_{bound} \ge$ 70
(from Ref. \cite{schuetti}).
        }
        \label{fig6}
\end{figure}

The almost identical behavior of the observed charge correlations
for different reactions, already suggested by the scatter plots,
is best appreciated when looking at the mean values
of these observables: 
In Fig. 5 the mean number of intermediate-mass
fragments is shown
as a function of $Z_{bound}$ for the reaction of $^{197}$Au on $^{197}$Au
at four bombarding energies. The rise and fall of fragment production
is seen to be independent of the projectile energy within the experimental
accuracy. This invariance also holds
for other charge correlations that have been found useful to
characterize the population of the partition space in the fragmentation
process \cite{schuetti,kreutz}.

\begin{figure}[tbh]
	\centerline{\epsfig{file=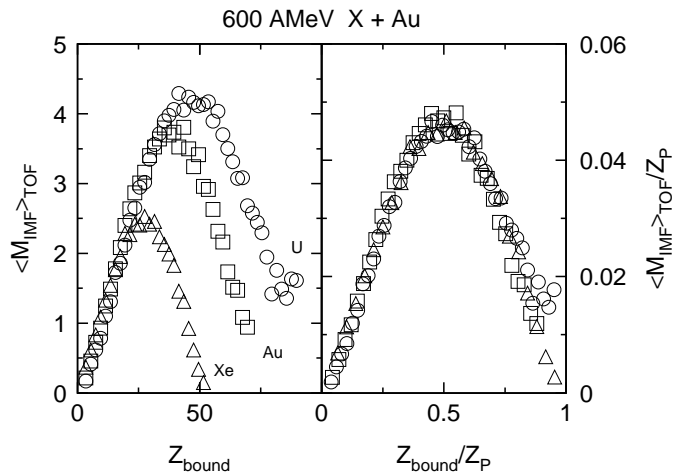,height=6.5cm}}
        \caption[]{\it\small
Left panel:
Mean multiplicity of intermediate-mass
fragments $\langle M_{IMF} \rangle_{TOF}$, observed with the TOF wall,
as a function of $Z_{bound}$ for the reactions
$^{238}$U on $^{197}$Au (circles), $^{197}$Au on $^{197}$Au (squares),
and $^{129}$Xe on $^{197}$Au (triangles) at $E/A$ = 600 MeV. Note that
also in $Z_{bound}$ only fragments detected with the TOF wall are included.

\noindent
Right panel:
The same data, as shown in the left panel, after normalizing both
quantities with respect to the
atomic number $Z_{P}$ of the projectile
(from Ref. \cite{schuetti}).
        }
        \label{fig7}
\end{figure}

The target invariance of the $M_{IMF}$ versus $Z_{bound}$ correlation
was first observed for collisions of $^{197}$Au projectiles
with C, Al, Cu, and Pb targets at 600 MeV per nucleon [23,25-27].
%\cite{hubele1,ogilvie,hubele2,kreutz}.
In Fig. 6 (top) the universal
nature of this correlation is demonstrated
for $^{238}$U projectiles at 1000 MeV per nucleon
and for a set of seven targets, ranging from Be to U.
The data for the lighter targets extend only over parts of the 
$Z_{bound}$ range. This is more clearly seen in the bottom part of the 
figure where the differential cross sections 
$d\sigma/dZ_{bound}$ for four out of the seven targets are shown.
From the cross sections,
by assuming a monotonic relation between $Z_{bound}$ and the
impact parameter, an empirical impact parameter scale was obtained.
Central collisions correspond to the smallest values of $Z_{bound}$
reached with a given target,
and given regions of $Z_{bound}$, 
in collisions with different targets,
correspond to different impact parameters. The cross sections were
found to depend somewhat on the bombarding energy. The range of
$Z_{bound}$ covered with, e.g., the Be or C targets increases with
increasing bombarding energy.

\begin{figure}[tbh]
	\centerline{\epsfig{file=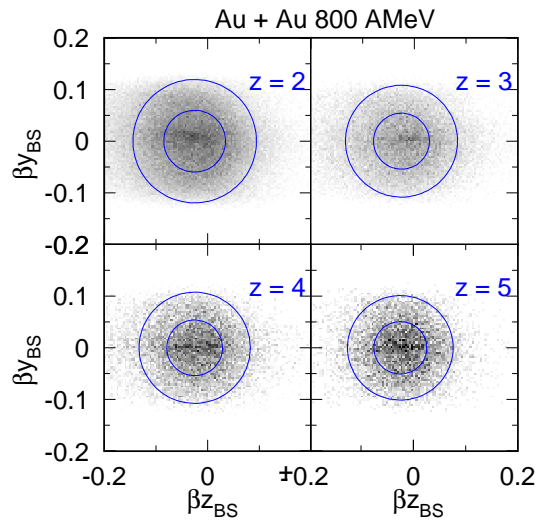,height=8cm}}
        \caption[]{\label{fig8} \it\small
Projections of the fragment velocity 
$\beta_{BS}$ in the moving frame (BS = beam system)
into the y-z-plane for products with atomic number $Z$ = 2 to 5
from the reaction $^{197}$Au on $^{197}$Au at $E/A$ = 800 MeV
(y and z denote the directions parallel to the magnetic field 
and along the beam direction, respectively).
The component $\beta y_{BS}$ is limited by the vertical acceptance
of the magnet, the circles are meant to guide the eye
(from Ref. \cite{schuetti1}).
        }
\end{figure}

The $\langle M_{IMF} \rangle$ versus $Z_{bound}$ correlation depends
on the mass of the projectile. The results obtained with the three
projectiles $^{129}$Xe, $^{197}$Au, and $^{238}$U
at 600 MeV per nucleon show that,
on the absolute scale, more fragments are produced in the decay of
heavier projectiles (Fig. 7, left-hand side).
However, a normalization with respect to the atomic number
$Z_P$ of the projectile
reduces the three curves to a single universal relation
(Fig. 7, right-hand side).

The observed $Z_{bound}$ scaling thus comprises the dependences
on the projectile and target mass and on the bombarding energy.
The data obtained at the AGS with beams of 10.6 GeV per nucleon,
in fact, indicate that it should be valid 
up to very high bombarding energies \cite{jain2,klmm2}. 
The reasons underlying this property of spectator decay may be sought in 
the mechanism of spectator excitation. Calculations with the intranuclear
cascade model \cite{yariv} 
suggest that the relation between the excitation energy
and the residual mass of the spectator should be 
universal \cite{schuetti,toneev}.
In this case, if the subsequent decay proceeds statistically,
the final fragmentation patterns will only depend on
$Z_{bound}$ and not on the particular entrance channel of the reaction.

\section{The equilibrated spectator source}
\label{Sec_4}

The observed universality of the spectator decay
suggests that a high degree of equilibrium is reached in the initial
stages of the reaction. This is confirmed by the analysis of the
kinetic variables in the moving frame of the spectator.

\begin{figure}[tbh]
	\centerline{\epsfig{file=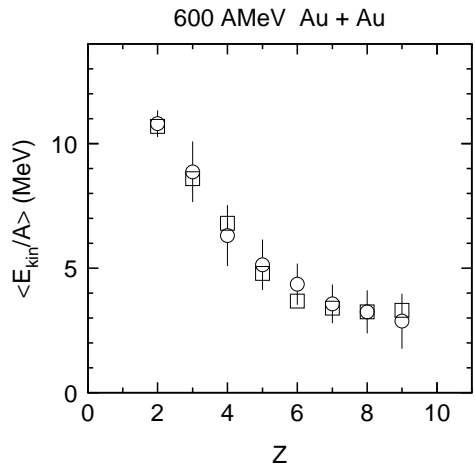,height=6.5cm}}
        \caption[]{\it\small
Mean kinetic energies per nucleon in the moving frame,
deduced from the transverse (circles) and longitudinal (squares)
momentum widths, for fragments from
the reaction $^{197}$Au on $^{197}$Au at $E/A$ = 600 MeV
and for 20 $\le Z_{bound} \le$ 60
(from Ref. \cite{schuetti}).
        }
        \label{fig9}
\end{figure}

The times and positions measured with the TOF wall were used 
to calculate the components of the fragment velocities in the reference
frame of the original projectile. Results obtained for light fragments
from the reaction
$^{197}$Au on $^{197}$Au at $E/A$ = 800 MeV are shown in Fig. 8.
In the y direction, the acceptance of the magnet limits
the observable range of velocities. Apart from this, the
distributions are seen to be isotropic to a very good approximation.
Gaussian widths fitted to the measured distributions of the velocity
components in the moving frame confirm this isotropy quantitatively. 

The intrinsic velocities have also been used to determine the fragment
kinetic energies in the frame of the decaying spectator.
Results for the reaction $^{197}$Au on
$^{197}$Au at 600 MeV per nucleon, integrated over
20 $\le Z_{bound} \le$ 60, are shown in Fig. 9. 
It was assumed that either
the longitudinal (squares) or the vertical transverse (circles)
degrees of freedom represent one third of the total kinetic
energies in the moving frame.
The agreement between the two sets of results reflects the
isotropy of the kinetic degrees of freedom.
The intrinsic kinetic energies do not depend on the bombarding
energy and thus 
are representative for the whole energy range over which the universal
spectator decay prevails.

The mean kinetic energies per unit fragment mass
$\langle E_{kin}/A \rangle$ decrease
rapidly with atomic
number $Z$.
In the limit of purely thermal contributions to the
kinetic energies, $\langle E_{kin}/A \rangle$ is expected to have
a 1/$A$ dependence
which is approximately observed. However, on the order of one half
of the kinetic energies in the rest frame of the decaying system may
originate from Coulomb repulsion and sequential decays of excited fragments
\cite{voli}. With this assumption the magnitude of the kinetic
temperature $T = 2/3 \cdot 1/2 \cdot \langle E_{kin} \rangle$ assumes
a value of approximately 15 MeV. This exceeds considerably
the emission temperatures $T \approx$ 5 MeV derived from the relative
isotopic abundances (see below) or from relative yields of particle
unbound states \cite{kunde91} which represents a well known but up to now 
not fully resolved problem [33-36]. 
%\cite{barz2,boal,barz3,bauer1}.

\section{Temperatures at breakup}
\label{Sec_5}

Several techniques have been developed for the measurement of
temperatures of excited nuclear systems \cite{morrissey}.
In the work leading to
the caloric curve of nuclei the method suggested
by Albergo {\it et al.} \cite{albergo} has been
used. It is based on the assumption of chemical equilibrium and
requires the measurement of double ratios of 
isotopic yields.

In the limit of thermal and chemical equilibrium,
the double ratio $R$ built from the yields $Y_i$ of two pairs of nuclides
with the same differences in neutron and proton numbers is given by 
\begin{equation}
R=\frac{Y_{1}/Y_{2}}{Y_{3}/Y_{4}} = 
a\cdot \exp(((B_{1}-B_{2})-(B_{3}-B_{4}))/T)
\label{EQ1}
\end{equation}
where $B_{i}$ denotes the binding energy of particle species i and the constant
$a$ contains their ground-state spins and mass numbers.
In order to make the ratios sufficiently sensitive to the 
temperature $T$ the double difference of binding energies
should be larger than the typical temperature to be measured. 
For this reason, $^{3}$He and $^{4}$He are a useful choice for forming 
one of the two ratios because the difference in binding energy is 20.6 MeV.
It may be combined with, e.g., the lithium yield ratio 
$^{6}$Li/$^{7}$Li or with the hydrogen yield ratios p/d or d/t. 
Mass spectra obtained for the
four isotopes $^{3}$He, $^{4}$He, $^{6}$Li, and$^{7}$Li
from the tracking analysis are shown in Fig. 10. The 
$^{3}$He yields reflect the sensitivity of this less strongly 
bound nuclide to the variation of the
temperature with impact parameter.

\begin{figure}[ttb]
	\centerline{\epsfig{file=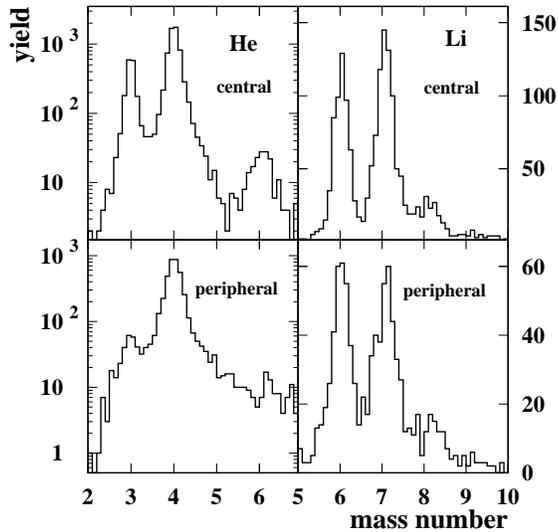,height=7.5cm}}
        \caption[]{\it\small
Mass spectra of He fragments (left panel)
and Li fragments (right panel)
for the reaction $^{197}$Au on $^{197}$Au at $E/A$ = 600 MeV.
The upper and lower panels
correspond to central and peripheral collisions, respectively
(from Ref. \cite{theo}).
        }
        \label{fig10}
\end{figure}

Solving Eq. (1) with
respect to $T$ yields, for the case of He and Li isotopes,
the following expression:
\begin{equation}
T_{HeLi,0} = 13.3 MeV/\ln(2.2\frac{Y_{^{6}Li}/Y_{^{7}Li}}{
Y_{^{3}He}/Y_{^{4}He}}).
\label{EQ2}
\end{equation}

The subscript 0 of $T_{HeLi,0}$ refers to the fact that
Eq. (1) is strictly valid only for the ground-state 
population of the considered isotopes at the breakup stage
which later may be modified by feeding from the
decay of excited states. 
The temperatures $T_{HeLi,0}$, therefore, will no longer be identical
to the breakup temperature.
The expected magnitude of this effect 
was investigated by performing
calculations with the quantum statistical model \cite{konopka}.
Results for an assumed
density $\rho /\rho_0$ = 0.3
are shown in Fig. 11 ($\rho_0$ denotes the saturation density of nuclei).
The relation between $T_{HeLi,0}$ 
and the input temperature of the model was found to be almost linear
which is also the case for, e.g., $T_{Hedt,0}$. Other temperature probes,
as illustrated for $T_{Hepd,0}$ in the figure, 
may be more strongly affected by sequential decays.
Variations of the input density within reasonable limits or
calculations with other decay models suggested that the accuracy of
these estimates may lie within $\pm$ 15\% \cite{theo}.
Based on these findings, the constant correction
$T_{HeLi} = 1.2 \cdot T_{HeLi,0}$, corresponding to the dotted line
in Fig. 11, was adopted \cite{pocho1}.

\begin{figure}[tbh]
	\centerline{\epsfig{file=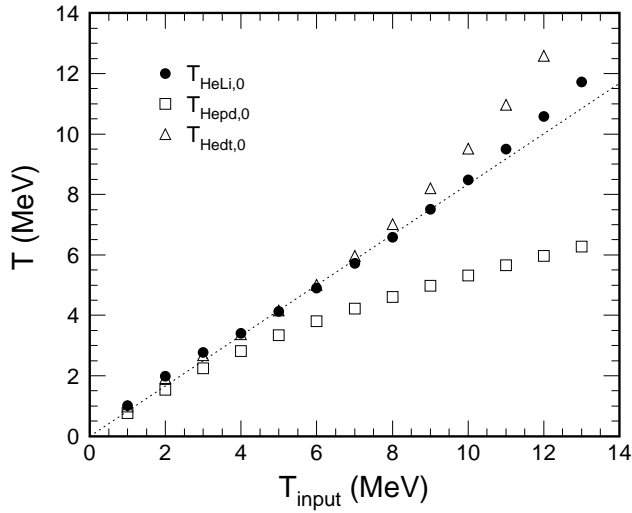,height=7cm}}
        \caption[]{\it\small
Temperatures $T_{HeLi,0}$, $T_{Hepd,0}$, and $T_{Hedt,0}$, according to the
quantum statistical model,
as a function of the input temperature $T_{input}$. A breakup
density $\rho/\rho_0$ = 0.3 is assumed. The dotted line represents the 
linear
relation $T_{input}$/1.2
(from Ref. \cite{hongfei}).
        }
        \label{fig11}
\end{figure}

Temperatures $T_{HeLi}$,
as deduced with this method from data measured in three experiments,
are shown in Fig. 12 as a function of $Z_{bound}$. 
Besides the results
obtained for the projectile decay in $^{197}$Au on $^{197}$Au collisions at
600 MeV and 1000 MeV per nucleon \cite{pocho1,theo}, 
also temperatures from a more recent
study of the target decay in the same reaction at 1000 MeV per nucleon
are given \cite{hongfei}.
In the latter experiment, $Z_{bound}$ was simultaneously measured for
the coincident projectile decay.
The temperatures increase slowly with decreasing $Z_{bound}$
in the range 20 $\le Z_{bound} \le$ 80 but then increase more quickly 
at small $Z_{bound}$ values. 

The agreement between the temperatures for the
projectile and the target spectators at 1000 MeV per nucleon
is expected from the symmetry of the reaction and 
illustrates the accuracy of the measurements.
The invariance with the bombarding energy, here established over the range
600 to 1000 MeV per nucleon, 
confirms the statistical interpretation of the observed universality 
of the spectator decay. 

\begin{figure}[tbh]
	\centerline{\epsfig{file=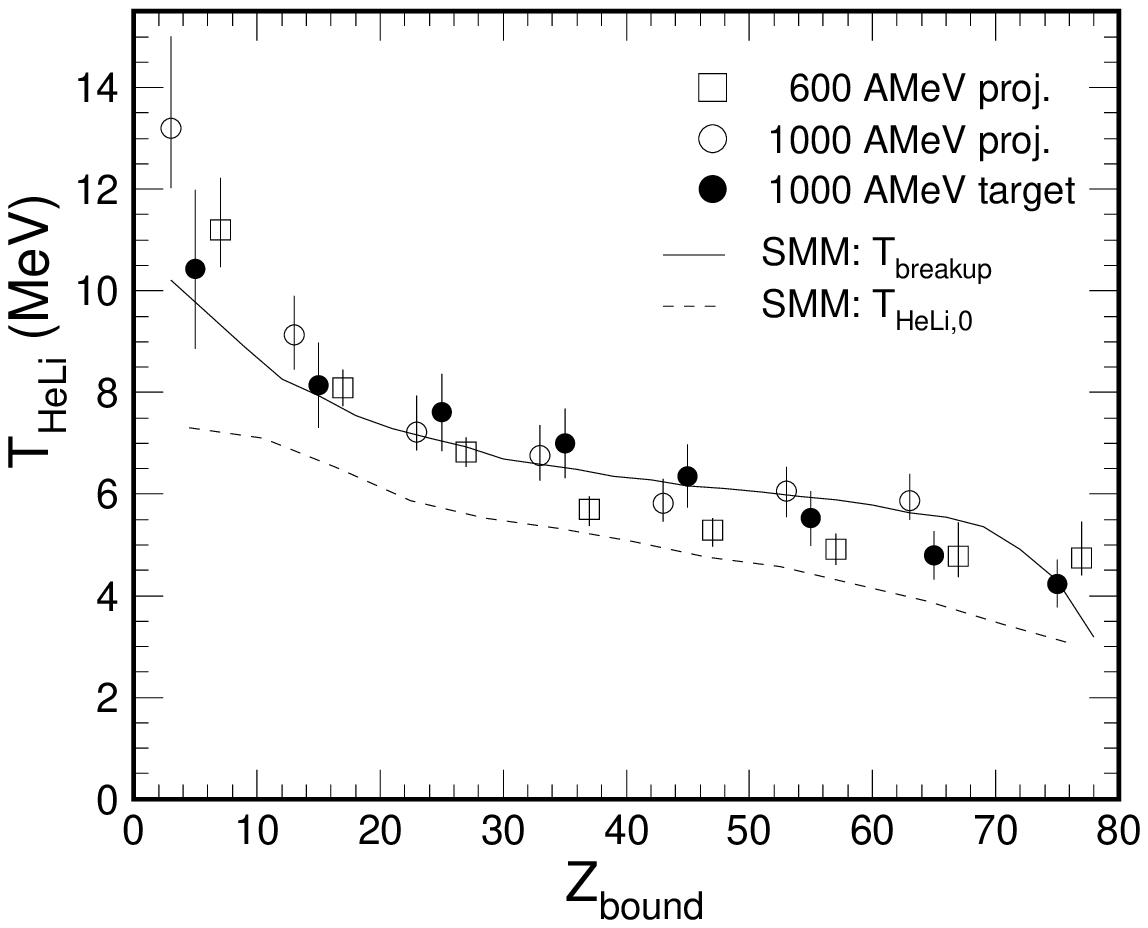,height=7cm}}
        \caption[]{\it\small
Temperatures $T_{HeLi}$ for target ($E/A$ = 1000 MeV) 
and projectile spectators ($E/A$ = 600 and 1000 MeV)
as a function of $Z_{bound}$. The data symbols represent averages over
bins of 10-units width. 
The full and dashed lines
represent the internal temperature $T_{breakup}$ and the uncorrected
isotopic temperature $T_{HeLi,0}$ as given by the calculations with the
statistical multifragmentation model
(from Ref. \cite{hongfei}).
        }
        \label{fig12}
\end{figure}

The lines shown in Fig. 12 represent results obtained with
the statistical multifragmentation model \cite{bondorf2}.
The excitation energy and mass of the ensemble of
excited spectator nuclei, required as input for the calculations,
were chosen in such a way 
that the correlation between the mean multiplicity 
of intermediate-mass fragments with
$Z_{bound}$ (Fig. 5) was well reproduced. 
Within the given experimental and methodical uncertainties, 
the resulting mean value of the breakup temperature (full line) is
in excellent agreement with the data.
This means that the description
of the fragmentation as a statistical process is internally
consistent in that the temperatures needed to reproduce the observed
partition patterns are equal to those measured. The dashed line gives
the uncorrected
isotopic temperature $T_{HeLi,0}$ deduced from the calculated
isotope yields. The difference to the breakup 
temperature represents the correction for secondary decay according
to the statistical multifragmentation model. It is in good qualitative
agreement with the adopted correction factor of 1.2.

The consequences of sidefeeding from higher lying states are
presently investigated
by several groups with different methods [42-47]. 
%\cite{tsang2,kolomiets,campi96,tsang3,xi,majka}.
The results differ considerably in magnitude as well as in the sign of the
required correction and, in some cases, exceed the $\pm$15\% margin
quoted above. Furthermore, temperatures for central collisions
at lower bombarding energies obtained from isotopic yield ratios and from
the population of particle-unstable resonances were found to deviate
in a systematic fashion from each other \cite{recent}.
These questions will have to be answered, eventually,
in order to maintain a quantitative level 
in the investigation of thermodynamic properties of excited nuclear systems. 

\section{Energy deposition}
\label{Sec_6}

Rather small fractions of the initial bombarding energy are imparted to the
spectator nuclei in relativistic collisions. The actual amounts
of energy deposition 
can only be reconstructed from the exit-channel configuration which
requires a complete knowledge of all decay products, including their
atomic numbers, masses, and kinetic energies.

A method to determine the excitation energy from the experimental data
along this line 
was first presented by Campi {\it et al.} \cite{campi94}
and applied to the earlier $^{197}$Au + Cu data \cite{kreutz}. 
The yields of hydrogen isotopes were determined
by extrapolating to $Z$ = 1 from the 
measured abundances for $Z \ge$ 2,
and the multiplicities of neutrons were inferred from a mass balance. 
The obtained asymptotic value of
$E_x/A$ = 23 MeV at $Z_{bound}$ = 0
is the sum of the binding energy of 8 MeV
and the kinetic energy of 15 MeV assigned to nucleons.

\begin{figure}[ttb]
	\centerline{\epsfig{file=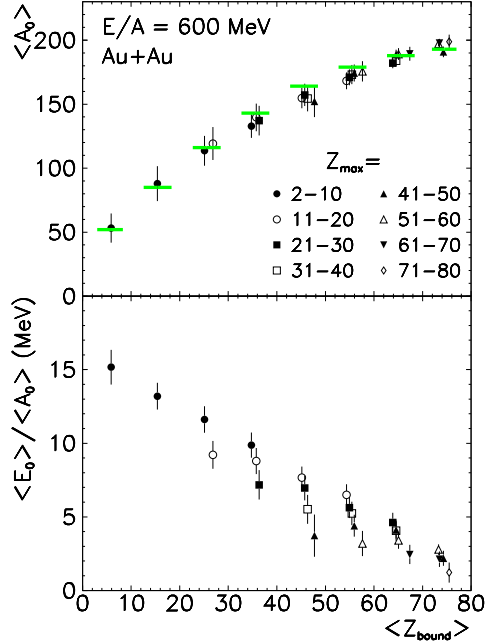,height=9cm}}
        \caption[]{\it\small
Reconstructed average mass $\langle A_0 \rangle$ (top) and excitation energy
$\langle E_0 \rangle / \langle A_0 \rangle$ (bottom) of the decaying
spectator system as functions of $Z_{bound}$ (abscissa) and of 
$Z_{max}$ (different data symbols). 
The horizontal bars represent the
masses according to the participant-spectator model at the empirical
impact parameter deduced from $d\sigma /dZ_{bound}$
(from Ref. \cite{pocho1}).
        }
        \label{fig13}
\end{figure}

In the same type of analysis with the more recent data
for $^{197}$Au + $^{197}$Au at 600 MeV per nucleon, the data on neutron
production measured with LAND were taken into account \cite{pocho1}. 
Since the hydrogen isotopes were not detected
assumptions concerning the overall $N/Z$ ratio of the spectator,
the intensity ratio of protons, deuterons, and tritons, and the
kinetic energies of hydrogen isotopes had to be made.
In addition, the EPAX parameterization \cite{suem90,botv95}
was used
for deriving masses from the atomic numbers of the detected fragments.
The uncertainties resulting from the variation of these quantities within
reasonable limits were included in the errors assigned to the results.

It is found that light particles and, in particular, the neutrons
contribute considerably to the total spectator energy $E_0$.
The balance of binding energies, i.e. the Q value associated with the
fragmentation process, amounts to about 40\% of it, fairly independent
of $Z_{bound}$.
The results for the mass $A_0$ and for the specific
excitation energy $E_0/A_0$
are given in Fig. 13. The data points represent the results for 10-unit-wide 
bins in a $Z_{max}$-versus-$Z_{bound}$ representation (cf. Fig. 4).
The mean mass $A_0$ decreases with decreasing $Z_{bound}$,
seems to be independent of $Z_{max}$,
and is in good agreement with the expectations from the geometric
participant-spectator model \cite{gosset}.
The smallest mean spectator
mass in the bin of $Z_{bound} \le$ 10 is $\langle A_0 \rangle \approx$ 50.
The excitation energy $E_0$ appears to be a function of both $Z_{bound}$ 
and $Z_{max}$; 
the higher values correspond to the smaller $Z_{max}$ values, i.e.
to more complete disintegrations of a system of given mass.
The maximum number of fragments, observed at $Z_{bound} \approx$ 40, is
associated with initial excitation energies of
$\langle E_0 \rangle/\langle A_0 \rangle \approx$ 8 MeV.
With decreasing $Z_{bound}$ the deduced excitation energies reach up
to $\langle E_0 \rangle/\langle A_0 \rangle \approx$ 16 MeV.

The experimentally determined energies fall in between
the higher predictions for the deposited energy of the intranuclear
cascade model and the much lower values obtained from analyses of
the final partitions with the statistical multifragmentation model
(Ref. \cite{schuetti} and references therein). A sequence of energies
with this ordering is not unreasonable in that the formation of the
equilibrated spectator in the initial reaction stages and its evolution
towards the final breakup stage may be accompanied by the emission
of fast light particles and, therefore, by a loss of excitation energy.
On the experimental side, it is presently investigated to what extent
the unexpectedly high kinetic energies of protons, preequilibrium 
emission, and 
collective phenomena, in particular the bounce-off of the
spectator systems, are influencing the
deduced energy deposits \cite{schuetti1}.
The result reported most recently \cite{schuetti} is about 15\% higher
than that shown in Fig. 13 which entered the caloric curve 
discussed in the next section.

\section{The caloric curve}
\label{Sec_7}

The pairwise correlation of the temperatures and excitation energies,
deduced as described in the last two sections,
results in the caloric curve shown in Fig. 14.
Besides the data from projectile decays following 
$^{197}$Au + $^{197}$Au collisions at 600 MeV per nucleon,
results from earlier experiments with $^{197}$Au targets at intermediate
energies 30 to 84 MeV per nucleon and for compound nuclei produced
in the $^{22}$Ne + $^{181}$Ta reaction are included \cite{trockel,borcea}. 
All temperatures were deduced following the same method.
For the reactions at intermediate energies, the excitation energy
of the target residues was obtained from an energy balance based
on moving-source analyses while, in the compound case, it is given by
the collision energy.

\begin{figure}[tbh]
	\centerline{\epsfig{file=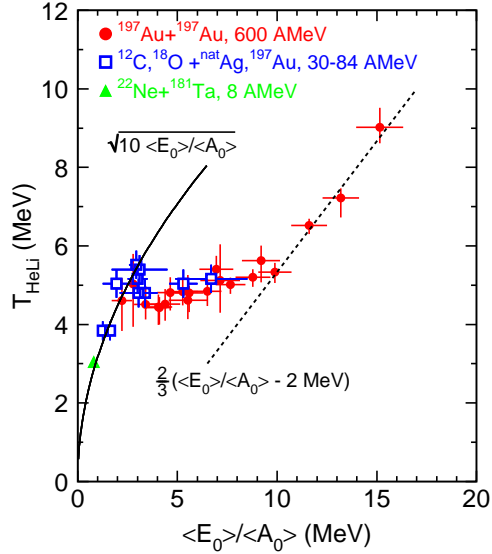,height=8cm}}
        \caption[]{\it\small
Caloric curve of nuclei as constituted by the
temperature T$_{HeLi}$
as a function of the excitation energy per nucleon.
The lines are explained in the text
(from Ref. \cite{pocho1}).
        }
        \label{fig14}
\end{figure}

One may first notice the consistency of the data obtained
from different types of reactions, suggesting that the smooth 
S-shaped curve may represent a more general property of excited nuclei.
In fact, at low energies, the deduced temperatures
$T_{HeLi}$ follow the low-temperature approximation for a Fermi-liquid,
confirmed by several studies in the fusion evaporation regime
\cite{nebbia,fabris}. The full line depicts this behavior for
a level density parameter $a = A/10$ MeV$^{-1}$.
At the high excitation energies, the rise of the temperature appears
to be linear with the excitation energy, with the slope of 2/3  
of a classical gas. 
In the limit of a free nucleon gas, the offset should be $\approx$ 8 MeV,
corresponding to the mean binding energy of nuclei.
A smaller offset may be caused by
a freeze-out at a finite density and by the finite fraction
of bound clusters and fragments of intermediate mass 
that are present even at these high excitation energies. The offset
of 2 MeV is consistent with a breakup density 
$\rho /\rho_0$ between 0.15 and 0.3 \cite{pocho1}.
A final assessment, however, will have to await the completion of the
ongoing analysis of the energy deposition.

Within the range of $\langle E_0\rangle /\langle A_0\rangle$ from
3 MeV to 10 MeV an almost constant value for $T_{HeLi}$ of about 4.5
to 5 MeV is observed. This plateau may be related to
the previous finding of almost constant emission temperatures
over a broad range of
incident energies which were deduced from the population of
particle unstable levels in He and Li fragments \cite{kunde91}.
The plateau is suggestive of a first-order phase transition with a
substantial latent heat (see also \cite{moretto96}). 
This is supported by the fact that
the increase in excitation energy is associated with a disintegration
into a larger number of fragments of smaller size as apparent from
Fig. 4 ($\langle E_0\rangle /\langle A_0\rangle$ has to be translated back
into the corresponding $Z_{bound}$ with the help of
Fig. 13). The surface energy
needed for the formation of smaller constituents limits the rise
of the temperature. This interpretation is consistent with the
good description of the plateau temperatures by 
the statistical multifragmentation model which is based on the droplet
model of nuclei (Fig. 12).

Alternative interpretations, in particular for the hitherto unobserved
temperature rise at the highest energies, have been presented by several
groups. The interpretation of 
Natowitz {\it et al.} \cite{natowitz}
relates the observed variation of the temperature to the variation of the
system mass (Fig. 13) and to the mass dependence of the limiting
temperatures obtained from theoretical descriptions of excited nuclei
at the limit of their stability \cite{bonche}.
In the expansion scenario modeled by Papp and N\"orenberg \cite{papp},
the temperatures
in the plateau region are found to be consistent with a spinodal
decomposition in the dynamically unstable region of the 
temperature-versus-density plane. The upbend at high excitation energies,
however, indicates a minimum breakup density rather than entry into
the vapor phase. The results were found to be sensitive to the
equation of state governing the expansion process which presents a
further motivation for aiming at a high level of accuracy in these
measurements.

\section{Critical features of multifragmentation}
\label{Sec_8}

The apparent signatures of a first-order phase transition in nuclei,
discussed in the last section, do not rule out the possibility that
critical phenomena may be observed. In finite systems, a second-order
phase transition is no longer characterized by a singular point,
the associated fluctuations are rather spread over a 
finite interval in temperature \cite{good84}.
Typical features of a first-order phase transition,
like a latent heat, and signals indicating the proximity of the
critical point, like diverging moments, are therefore not necessarily 
inconsistent.

The observation of a power law dependence of the fragment mass yields 
in reactions of energetic protons 
with Kr and Xe targets \cite{finn,hirsch}, and its association with Fisher's
prediction for droplet formation at the critical point \cite{fisher}
has initiated
the intensive search for signatures of criticality 
\cite{huefner,acicast,richert}. 
The systematic investigations showed that the power 
law exponent $\tau$ approaches a value of $\approx$ 2.5 at high
bombarding energies \cite{traut}. This is consistent with the 
limits 2.0 $\le \tau \le$ 3.0 given by the theory of critical 
phenomena \cite{stauffer}. While the interpretation of inclusive
mass spectra was criticized \cite{aich88}, it was shown in exclusive
measurements that the mass or charge distributions may approach
the pure power law for certain values of the chosen sorting variable,
e.g. $Z_{bound}$.
In this case, the sorting variable may serve as the parameter
controlling the distance to the critical point or critical region.
For the fragmentation of $^{197}$Au projectiles at 1~GeV per
nucleon, this value is $Z_{bound} \approx$ 35 \cite{woerner} which
is close to the maximum fragment production (Fig. 5).
Examples of $Z_{bound}$-gated spectra are shown in Ref. \cite{kreutz}. 

A more recent, equally stimulating, observation was the similarity
of the charge fluctuations in fragmentation with those given
by three-dimensional percolation on a small lattice \cite{campi88}.
Percolation is a mathematical model exhibiting a 
second-order phase transition 
in the limit of infinite lattice size \cite{stauffer}.
Campi, therefore, concluded that 'nuclei break up like finite systems that 
show a clean phase transition in infinite size'.
In the meanwhile, it has been demonstrated that 
nearly perfect descriptions of the partitions 
observed in multifragmentation reactions may be obtained with the
percolation model \cite{kreutz}. 

The capability of percolation to describe the partition space 
is not trivial but it is also 
not unique to percolation. The fragment-charge correlations
from the first experiments have been reproduced to high
accuracy with the statistical multifragmentation model,
but also with a variety of other models, such as
classical-cluster formation
\cite{garcia}, fragmentation-inactivation binary \cite{botet},
and restructured-aggregation models \cite{leray}.
Some of these models exhibit
critical behavior in the limit of an infinite number of constituents.
The apparent universality, inherent to disordered systems of
interacting nucleons and to the models describing their fragmenting
so well, remains to be understood \cite{elat95}.

\begin{figure}[tbh]
	\centerline{\epsfig{file=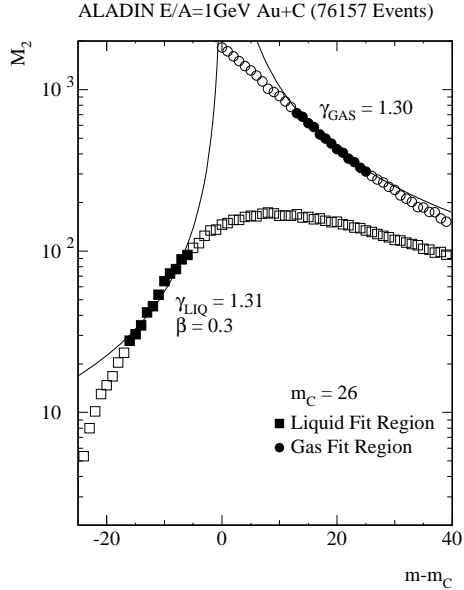,height=8cm}}
        \caption[]{\it\small
Second moment $M_2$ of the fragment-charge distribution, with (circles) and
without (squares) including the largest fragment, as a function of
the charged particle multiplicity $m$ 
for $^{197}$Au + C at $E/A$ = 1 GeV. The assumed critical multiplicity
is $m_c$ = 26, the full symbols indicate the selected fit regions
(from Ref. \cite{woerner}).
        }
        \label{fig15}
\end{figure}

A further step was taken by the EOS collaboration who have reported
values for critical-point exponents from 
the charge correlations measured for the
$^{197}$Au on C reaction at 1~GeV per nucleon 
\cite{gilkes,hauger,elliott96}.
The analysis takes extensive recourse to results obtained for
percolation. Below the critical point,
the largest cluster of the partition is associated with the
percolating cluster which corresponds to the 
liquid phase in a liquid-gas transition.
It grows in proportion to $t^{\beta}$ with the relative distance
$t$ from the critical point. The role of the susceptibility
in magnetic systems or the isothermal compressibility in
a liquid, i.e. the response to the external field, 
is assumed by the second moment $M_2$
of the fragment size distributions. It diverges with the
power of -$\gamma$ as the critical point is approached (both $\beta$
and $\gamma$ are positive numbers).

With the guidance provided by percolation studies on small lattices,
recipes were developed on how to extract critical exponents
from the data \cite{elliott94}. 
The reported results $\beta = 0.29 \pm 0.02$ and 
$\gamma = 1.4 \pm 0.1$ are close to those of a liquid-gas system
and significantly different from those of percolation or mean-field
theory \cite{gilkes}. This conclusion relies on the correct
assessment of the systematic errors inherent to the procedure
which, however, has been questioned by other authors 
\cite{bauer95,bauer95a}.
A critical discussion may also be found in the contributions
to Refs. \cite{bormio95,acicast} by M\"uller {\it et al.} who, besides
percolation, have studied Ising models on small lattices.

A critical-exponent analysis has been carried out with the 
ALADIN data for $^{197}$Au and $^{238}$U fragmentations \cite{woerner}.
The same numerical results were obtained if the procedure 
of Ref. \cite{gilkes} was followed in detail.
However, also the same difficulties arising from
the finite system size were encountered. As an example, the
fits used to extract the exponent $\gamma$ are shown in Fig. 15. 
The figure is virtually identical to the corresponding figure 
constructed from the EOS data \cite{ritter}. 
Potential divergences at the critical point that will appear in
the infinite system are smoothed over finite regions of the chosen
control parameter, the charged particle multiplicity $m$. 
The choice of the fit regions where the data are assumed to
reflect the critical behavior is crucial. It has to rely on methods obtained
from the study of other systems which, at this time, are not yet
proven unambiguous. A consistent method was found for
determining the exponent $\tau$ of the power law obeyed by the element
distribution \cite{woerner}. This may be related to the fact that
$\tau$ describes a behavior {\it at} the critical point, and not the
behavior how the critical point is approached. The finite size
of the system may be less crucial in this case. 

\section{Conclusions and perspectives}
\label{Sec_10}

The systematic set of data now available 
reveals the universal nature of multi-fragment
decays of excited spectator nuclei at relativistic energies.
It suggests that the correlation of excitation energy and mass
of the produced spectator systems and the statistical nature of
their decay are independent of the specific entrance channel.
The emulsion data as well as the insight provided by calculations
indicate that it should prevail up to very high
bombarding energies with virtually unchanged features. There are lower
limits of the bombarding energy below which the spectator excitation
does not suffice to induce a complete
disassembly with maximum fragment production. These threshold
energies depend on the collision partner, as shown for the case of
gold nuclei in Fig. 16. Besides data measured with the ALADIN
spectrometer in inverse kinematics, 
also data for $^{197}$Au on $^{197}$Au
at lower energies \cite{tsang93}, for $^{84}$Kr on $^{197}$Au 
\cite{peaslee}, and for $^{4}$He on $^{197}$Au \cite{lips} are
included. The hatched line indicates these threshold energies as
a function of the target mass. If the bombarding energy is raised
above the threshold,
the maximum fragment production will shift to more peripheral collisions.
Below the threshold, the production of heavy residues will be 
the dominating process.

\begin{figure}[tbh]
	\centerline{\epsfig{file=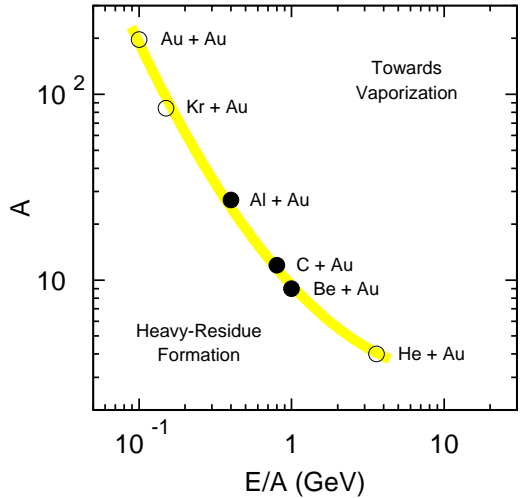,height=7cm}}
        \caption[]{\it\small
Mass number $A$ of the collision partner versus bombarding energy
for reactions of $^{197}$Au nuclei for which maximum fragment production 
has been observed in central collisions
(from Ref. \cite{schuetti}).
        }
        \label{fig16}
\end{figure}

As one moves along the hatched line towards the central collisions
of heavy systems, dynamical phenomena become important.
The largest fragment multiplicities 
measured so far were observed in central
$^{197}$Au on $^{197}$Au collisions at a bombarding energy of
100 MeV per nucleon \cite{tsang93}.
The analysis of the kinetic 
energy spectra in these reactions has revealed a
considerable collective outward motion (radial flow),
superimposed on the random motion
of the constituents at the breakup stage \cite{hsi}.
The associated collective energy constitutes
up to one-half of the incident kinetic energy in
the center-of-mass frame and increases approximately linearly over
the range of bombarding
energies up to 1000 MeV per nucleon \cite{reisdorf}. 
Furthermore, the fragment formation was found to be sensitive to the flow
dynamics. The measured element yields 
are systematically correlated with the magnitude of the observed
radial flow \cite{reisdorf,kunde95}. 

This excursion to central collisions, not covered in the
main part of these notes, is meant to demonstrate that
heavy ion reactions at relativistic energies offer wide
possibilities to study the response of excited nuclear matter under
different conditions. The interplay of dynamical
and statistical effects in these violent central collisions is of
high current interest \cite{bond94}. The measurement of 
breakup temperatures for these reactions, where equilibrium
may be established only locally, has already been started 
and promises to produce stimulating new results \cite{recent}.

Large dynamical effects are not expected in light-particle induced
reactions with heavy targets, i.e. at the lower end of the hatched line in
Fig. 16. Here the main problem arises from the fact that the cross sections
for large energy transfers to the target with maximum fragment 
production are comparatively small 
\cite{lips, kwiat,pienk}.
New experimental results including exclusive measurements with beams
of antiprotons \cite{gold96} and of protons with energies up to 14 GeV
\cite{viola96} have been obtained very recently, and a comprehensive
picture of the light-ion reactions should emerge rather soon.

The phenomenology of multifragment-decays, in particular the
partitioning of the decaying systems, has been established
with high accuracy, as shown in the first part of these notes.
The critical discussion of the new methods for measuring
thermodynamic properties of excited nuclear systems is important.
Here the systematic errors will have to be further reduced, which seems 
feasible with improved experiments and model calculations. 
The search for critical phenomena has to take  
the small number of constituents into account, and
concepts developed for macroscopic systems have to be applied with caution.
The finite system size, on the other hand, offers wide possibilities
for model studies based on different approaches, aiming at an
understanding of the universal properties of
nuclear fragmentation and of fragmentation and clustering
phenomena in other fields. 
\vspace{0.2cm}

{\it I am highly indebted to my colleagues of the ALADIN collaboration
for valuable discussions and support during the preparation of this 
manuscript.}

\vspace{0.8cm}

\renewcommand{\baselinestretch}{0.85}
\Large                % 1. Huerde:
                      % Erst groesser, um dann normal mit kleinerem
\normalsize                % Zeilenabstand zu schreiben


\begin{thebibliography}{99}


\bibitem{siemens}
P.J. Siemens, 
{\it Nature (London)} 305: 410 (1983).

\bibitem{huefner}
J.~H\"ufner, 
{\it Phys. Rep.} 125: 129 (1985).

\bibitem{gross1}
D.H.E.~Gross, 
{\it Rep.~Prog.~Phys.} 53: 605 (1990).

\bibitem{bondorf2}
J.P.~Bondorf {\it et al.},
{\it Phys Rep.} 257: 133 (1995).

\bibitem{jaqaman}
H.~Jaqaman {\it et al.},
\PR{C 27}{1983}{2782}.

\bibitem{brack}
M.~Brack {\it et al.}, 
{\it Phys. Rep.} 123: 275 (1985).

\bibitem{bertsch}
G. Bertsch and P.J. Siemens, 
\PL{126 B}{1983}{9}.

\bibitem{stocker}
W.~Stocker,
\PL{142 B}{1984}{319}.

\bibitem{pocho1}
J.~Pochodzalla {\it et al.},
\PRL{75}{1995}{1040}.

\bibitem{gilkes}
M.L.~Gilkes {\it et al.},
\PRL{73}{1994}{1590}.

\bibitem{schuetti}
A.~Sch\"uttauf {\it et al.},
\NP{A 607}{1996}{457}.

\bibitem{recent}
for the most recent reports of the collaboration see, e.g.,\\
G.~Imm\'e {\it et al.},
to appear in {\it Proceedings of the 1st Catania Relativistic Ion Studies:
Critical Phenomena and Collective Observables}, Acicastello, Italy, 1996;\\
W.F.J.~M\"{u}ller {\it et al.}, ibid.;\\
J.~Pochodzalla {\it et al.}, ibid.

\bibitem{hirschegg}
{\it Proceedings of the International Workshop XXII}, 
Hirschegg, 1994, edited by
H. Feldmeier and W. N\"orenberg (GSI, Darmstadt, 1994).

\bibitem{bormio95}
{\it Proceedings of the XXXIII International Winter Meeting on
Nuclear Physics}, Bormio, 1995,
edited by I. Iori (Ricerca Scientifica ed Educazione Permanente,
Milano, 1995).

\bibitem{snowbird}
{\it Proceedings of the 12th Winter Workshop on Nuclear Dynamics},
Snowbird, Utah, USA, 1996.

\bibitem{acicast}
{\it Proceedings of the 1st Catania Relativistic Ion Studies:
Critical Phenomena and Collective Observables}, Acicastello, Italy, 1996.

\bibitem{moretto}
For a recent review see L.G.~Moretto and G.J. Wozniak,
{\it Ann. Rev. Nucl. Part. Science} 43: 379 (1993).

\bibitem{jakob}
B.~Jakobsson {\it et al.},
\ZP{A 307}{1982}{293}.

\bibitem{fried}
E.M.~Friedlander {\it et al.},
\PR{C 27}{1983}{2436}.

\bibitem{jain2}
P.L. Jain {\it et al.},
\PR{C 50}{1994}{1085}.

\bibitem{klmm2}
M.L.~Cherry {\it et al.},
\PR{C 52}{1995}{2652}.

\bibitem{rusch}
G.~Rusch {\it et al.},
\PR{C 49}{1994}{901}.

\bibitem{hubele1}
J.~Hubele {\it et al.},
\ZP{A 340}{1991}{263}.

\bibitem{kunze}
W.D. Kunze,
PhD thesis, Universit\"at Frankfurt, 1996, unpublished.

\bibitem{ogilvie}
C.A.~Ogilvie {\it et al.}, 
\PRL{67}{1991}{1214}.

\bibitem{hubele2}
J.~Hubele {\it et al.},
\PR{C 46}{1992}{R1577}.

\bibitem{kreutz}
P.~Kreutz {\it et al.},
\NP{A 556}{1993}{672}.

\bibitem{yariv}
Y.~Yariv and Z.~Fraenkel,
\PR{C 20}{1979}{2227};
\PR{C 24}{1981}{488}.

\bibitem{toneev}
V.S.~Barashenkov and V.D.~Toneev,
Interactions of high-energy particles and atomic nuclei with nuclei,
Moscow, Atomizdat, 1972 (in Russian);\\
V.D.~Toneev, private communication (1994).

\bibitem{schuetti1}
A.~Sch\"uttauf,
PhD thesis, Universit\"at Frankfurt, 1996, unpublished.

\bibitem{voli}
V.~Lindenstruth, PhD thesis, Universit\"at Frankfurt, 1993, 
report GSI-93-18.

\bibitem{kunde91}
G.J.~Kunde {\it et al.},
\PL{B 272}{1991}{202}.

\bibitem{barz2}
H.W. Barz {\it et al.},
\PL{B 217}{1989}{397}.

\bibitem{boal}
D.H. Boal {\it et al.},
\PRL{62}{1989}{737}.

\bibitem{barz3}
H.W. Barz {\it et al.},
\PL{B 228}{1989}{453}.

\bibitem{bauer1}
W. Bauer,
\PR{C 51}{1995}{803}.

\bibitem{morrissey}
D. Morrissey {\it et al.},
{\it Ann. Rev. Nucl. Part. Science} 44: 27 (1994).

\bibitem{albergo}
S.~Albergo {\it et al.},
{\it Il Nuovo Cimento} 89 A: 1 (1985).

\bibitem{konopka}
J. Konopka {\it et al.},
\PR{C 50}{1994}{2085}.

\bibitem{theo}
T.~M\"ohlenkamp, 
PhD thesis, Universit\"at Dresden, 1996, unpublished.

\bibitem{hongfei}
Hongfei Xi {\it et al.},
to be published.

\bibitem{tsang2}
M.B.~Tsang {\it et al.},
\PR{C 53}{1996}{R1057}.

\bibitem{kolomiets}
A.~Kolomiets {\it et al.},
\PR{C 54}{1996}{R472}.

\bibitem{campi96}
X.~Campi {\it et al.},
{\it Phys. Lett.} B, in print.

\bibitem{tsang3}
M.B.~Tsang {\it et al.},
preprint MSUCL-1035 (1996).

\bibitem{xi}
Hongfei Xi {\it et al.},
preprint MSUCL-1040 (1996).

\bibitem{majka}
Z.~Majka {\it et al.},
preprint 96-03, Texas A\&M University (1996).

\bibitem{campi94}
X.~Campi {\it et al.},
\PR{C 50}{1994}{R2680}.

\bibitem{suem90}
K.~S\"ummerer {\it et al.},
\PR{C 42}{1990}{2546}.

\bibitem{botv95}
A.S.~Botvina {\it et al.},
\NP{A 584}{1995}{737}.

\bibitem{gosset}
J.~Gosset {\it et al.},
\PR{C 16}{1977}{629}.

\bibitem{trockel}
R.~Trockel {\it et al.},
\PR{C 39}{1989}{729};
R.~Trockel,
PhD thesis, Universit\"at Heidelberg, 1988, unpublished.

\bibitem{borcea}
C.~Borcea {\it et al.},
\NP{A 415}{1984}{169}.

\bibitem{nebbia}
G.~Nebbia {\it et al.},
\PL{B 176}{1986}{20}.

\bibitem{fabris}
D.~Fabris {\it et al.},
\PL{B 196}{1987}{429}.

\bibitem{moretto96}
L.G.~Moretto {\it et al.},
\PRL{76}{1996}{2822}.

\bibitem{natowitz}
J.B.~Natowitz {\it et al.},
\PR{C 52}{1995}{R2322}.

\bibitem{bonche}
P.~Bonche {\it et al.},
\NP{A 436}{1986}{265}.

\bibitem{papp}
G.~Papp and W.~N\"orenberg,
preprint GSI-95-30 (1995).

\bibitem{good84}
A.L.~Goodman {\it et al.},
\PR{C 30}{1984}{851}.

\bibitem{finn}
J.E.~Finn {\it et al.},
\PRL{49}{1982}{1321}.

\bibitem{hirsch}
A.S.~Hirsch {\it et al.},
\PR{C 29}{1984}{508}.

\bibitem{fisher}
M.E.~Fisher,
{\it Physics (N.Y.)} 3: 255 (1967).

\bibitem{richert}
J. Richert,
{\it Int. J. Mod. Phys.} E2: 679 (1993).

\bibitem{traut}
W.~Trautmann {\it et al.},
\ZP{A 344}{1993}{447}.

\bibitem{stauffer}
D.~Stauffer and A.~Aharony,
{\it Introduction to Percolation Theory},
Taylor \& Francis, London (1992).

\bibitem{aich88}
J.~Aichelin {\it et al.},
\PR{C 37}{1988}{2451}.

\bibitem{woerner}
A. W\"orner,
PhD thesis, Universit\"at Frankfurt, 1995, unpublished.

\bibitem{campi88}
X.~Campi,
\PL{B 208}{1988}{351}.

\bibitem{garcia}
J.B. Garcia and C. Cerruti,
\NP{A 578}{1994}{597}.

\bibitem{botet}
R. Botet and M. Ploszajczak,
\PL{B 312}{1993}{30};
{\it Acta Physica Polonica} B25: 353 (1994).

\bibitem{leray}
S. Leray and S. Souza,
{\it Proceedings of Second European Biennial Conference
on Nuclear Physics}, Meg\`{e}ve 1993, edited by D. Guinet
(World Scientific, Singapore, 1995) p. 81.

\bibitem{elat95}
B. Elattari {\it et al.},
\PL{B 356}{1995}{181};
\NP{A 592}{1995}{385}.

\bibitem{hauger}
J.A.~Hauger {\it et al.},
\PRL{77}{1996}{235}.

\bibitem{elliott96}
J.B.~Elliott {\it et al.},
\PL{B 381}{1996}{35}.

\bibitem{elliott94}
J.B.~Elliott {\it et al.},
\PR{C 49}{1994}{3185}.

\bibitem{bauer95}
W.~Bauer and W.A.~Friedman,
\PRL{75}{1995}{767}.

\bibitem{bauer95a}
W.~Bauer and A.S.~Botvina,
\PR{C 52}{1995}{R1760}.

\bibitem{ritter}
H.G.~Ritter {\it et al.},
\NP{A 583}{1995}{491c}.

\bibitem{tsang93}
M.B.~Tsang {\it et al.},
\PRL{71}{1993}{1502}.

\bibitem{peaslee}
G.F.~Peaslee {\it et al.},
\PR{C 49}{1994}{R2271}.

\bibitem{lips}
V.~Lips {\it et al.},
\PRL{72}{1994}{1604}.

\bibitem{hsi}
W.C. Hsi {\it et al.},
\PRL{73}{1994}{3367}.

\bibitem{reisdorf}
W. Reisdorf {\it et al.},
in Ref. \cite{hirschegg}, p.93.

\bibitem{kunde95}
G.J. Kunde {\it et al.},
\PRL{74}{1995}{38}.

\bibitem{bond94}
J. Bondorf {\it et al.},
\PRL{73}{1994}{628}.

\bibitem{kwiat}
K.~Kwiatkowski {\it et al.},
\PRL{74}{1995}{3756}.

\bibitem{pienk}
L.~Pienkowski {\it et al.},
\PL{B 336}{1994}{147}.

\bibitem{gold96}
F.~Goldenbaum {\it et al.},
\PRL{77}{1996}{1230}.

\bibitem{viola96}
V.E.~Viola {\it et al.},
{\it Proceedings of Third International Conference
on Nuclear Physics at Storage Rings}, Bernkastel-Kues, Germany, 1996.

\end{thebibliography}
\end{document}